\documentclass[12pt]{article}

\newtheorem{thm}{Theorem}[section]

\font\myfont=msbm10 scaled \magstep1

\def\ZZ{\hbox{\myfont\char'132}}
\def\NN{\hbox{\myfont\char'116}}

\begin{document}

\begin{center}
{\Large On  Darboux Integrable Semi-Discrete Chains}

\vskip 0.2cm

{Ismagil Habibullin}\footnote{e-mail: habibullinismagil@gmail.com}\\

{Ufa Institute of Mathematics, Russian Academy of Science,\\
Chernyshevskii Str., 112, Ufa, 450077, Russia}\\
\bigskip

{Natalya Zheltukhina}\footnote{e-mail: natalya@fen.bilkent.edu.tr}

{Department of Mathematics, Faculty of Science,
 \\Bilkent University, 06800, Ankara, Turkey \\}

\bigskip

{Alfia Sakieva}\footnote{e-mail: alfiya85.85@mail.ru}

{Ufa Institute of Mathematics, Russian Academy of Science,\\
Chernyshevskii Str., 112, Ufa, 450077, Russia}\\

\end{center}

\begin{abstract}
Differential-difference equation
$\frac{d}{dx}t(n+1,x)=f(x,t(n,x),t(n+1,x),\frac{d}{dx}t(n,x))$
with unknown $t(n,x)$ depending on continuous  and discrete
variables $x$ and  $n$ is studied. We call an equation of such kind Darboux
integrable, if there exist two functions (called integrals) $F$  and $I$ of a finite
number of dynamical variables
such that $D_xF=0$ and $DI=I$, where $D_x$ is the operator of
total differentiation with respect to $x$, and
$D$ is the shift operator: $Dp(n)=p(n+1)$. It is proved that the integrals can be
brought to some canonical form. A method of construction of an explicit formula for general solution
to Darboux integrable chains is discussed and for a class of chains such solutions are found.
\end{abstract}


\section{Introduction}

In this paper we  study   Darboux integrable
semi-discrete chains of the form
\begin{equation}\label{dhyp}
\frac{d}{dx}t(n+1, x)=f(x,t(n, x),t(n+1, x),\frac{d}{dx}t(n, x))\, .
\end{equation}
Here unknown function $t=t(n,x)$ depends on discrete and continuous variables $n$ and $x$ respectively;
function $f=f(x,t,t_1,t_x)$ is assumed to
be locally analytic, and $\frac{\partial f}{\partial t_x}$ is not
identically zero. The last two decades the discrete phenomena have become very popular due
to various important applications (for more details see \cite{Zabrodin}-\cite{GKP} and references therein).
%Let us begin with the rigorous definitions.

%\subsection{Darboux integrability definition for semi-discrete chains}

Below we use a subindex to indicate the shift of the discrete
argument: $t_k=t(n+k,x)$, $k\in \ZZ$,   and derivatives with respect
to $x$: $t_{[1]}=t_x=\displaystyle{\frac{d}{d x}}t(n,x),$
$t_{[2]}=t_{xx}=\displaystyle{\frac{d^2}{d x^2}}t(n,x)$,
$t_{[m]}=\frac{d^m}{dx^m}t(n,x)$, $m\in \NN$.  Introduce {\it the set of
dynamical variables} containing $\{t_k\}_{k=-\infty}^{\infty};$
$\{t_{[m]}\}_{m=1}^{\infty}$.
 We denote through $D$ and $D_x$ the shift operator and the
operator of the total derivative with respect to $x$ correspondingly. For
instance, $Dh(n,x)=h(n+1,x)$ and
$D_xh(n,x)=\frac{d}{d x}h(n,x)$.

Functions $I$ and $F$, both depending on $x$, $n$,  and a finite number of
dynamical variables, are called  respectively {\it $n$- and $x$-integrals}
of (\ref{dhyp}), if  $DI=I$ and $D_xF=0$ (see also
\cite{AdlerStartsev}).
Clearly, any  function depending on $n$ only, is an $x$-integral, and any
function, depending on $x$ only, is an $n$-integral. Such integrals are called {\it trivial}
integrals. One can see that any $n$-integral $I$ does
not depend on variables $t_{m}$, $m\in \ZZ\backslash\{0\}$,
 and any $x$-integral $F$
does not depend on variables $t_{[m]}$, $m\in \NN$.

Chain (\ref{dhyp}) is called {\it Darboux integrable}
if it admits a nontrivial $n$-integral  and a nontrivial
$x$-integral.

The basic ideas on integration of partial differential equations of the hyperbolic type
go back to classical works by Laplace, Darboux, Goursat, Vessiot, Monge, Ampere, Legendre,
Egorov, etc. Notice that understanding of integration as finding an explicit formula for a
general solution was later replaced by other, in a sense less obligatory, definitions.
For instance, the Darboux method for integration of hyperbolic type equations consists of
searching for integrals in   both directions followed by the reduction of the equation to
two ordinary differential equations. In order to find integrals, provided that they exist,
Darboux used the Laplace cascade method.
An alternative, more algebraic approach based on the characteristic vector fields was used by
Goursat and Vessiot. Namely this method allowed Goursat to get a list of integrable equations
\cite{GOURSAT}. An important contribution to the development of the algebraic method
investigating  Darboux integrable equations was made by A.B.~Shabat who introduced the
notion of the characteristic algebra of the hyperbolic equation
\begin{equation}\label{hyp}
u_{x,y}=f(x,y,u,u_x,u_y)\, .
\end{equation}
It turned out that the operator $D_y$ of total differentiation, with
respect to the variable $y$, defines a derivative in the
characteristic algebra in the direction of  $x$. Moreover, the
operator $ad_{D_y}$ defined according to the rule
$ad_{D_y}X=[D_y,X]$ acts on the generators of the algebra in a very
simple way. This makes it possible to obtain effective integrability
conditions for the equation (\ref{hyp}) (see \cite{ShabatYamilov}).

A.V. Zhiber and F.Kh. Mukminov investigated the structure of the characteristic
algebras for the so-called quadratic systems containing the Liouville equation
and the sine-Gordon equation (see \cite{ZhiberMukminov}). In \cite{ZhiberMukminov} and
\cite{ZhiberMurtazina} the very nontrivial connection between characteristic algebras
and Lax pairs of the hyperbolic S-integrable equations and systems of equations is studied,
and perspectives on the application of the characteristic algebras to classify such kinds of
equations are discussed.

Recently the concept of the characteristic algebras has been
defined for discrete models. In our articles \cite{hp}-\cite{HZhP2}
an effective algorithm was worked out to classify Darboux integrable
models. By using this algorithm some new classification results were
obtained. In
\cite{h2010} a method of classification of S-integrable discrete
models is suggested based on the concept of characteristic algebra.

The article is organized as follows. In Section 2 characteristic algebras are defined for the chain (\ref{dhyp}). In Section 3 we describe the structure of $n$-integrals and $x$-integrals of the Darboux integrable chains of general
form (\ref{dhyp}) (see Theorems \ref{Thm1} and \ref{Thm2}). Then we show that one can
choose the minimal order $n$-integral and the minimal order $x$-integral of a special
canonical form, important for the purpose of classification (see Theorems
 \ref{Thm3} and \ref{Thm4}).

The complete classification of a particular case
$t_{1x}=t_x+d(t_1,t)$ in \cite{HZhP2} was done due to the finiteness
of the characteristic algebras in both directions. However,
algebras themselves were not described. In Subsections 4.1 and 4.2
we fill up this gap and represent the tables of multiplications
for all of these algebras.

The problem of finding explicit solutions for Darboux integrable
models is rather complicated. Even in the mostly studied case of PDE
$u_{xy}=f(x,y,u,u_x,u_y)$ this problem is not completely solved. In
Subsection 4.3 we give the explicit formulas for general solutions
of the integrable chains in  the particular case
$t_{1x}=t_x+d(t,t_1)$ (see Theorem \ref{TheoremExplicit}).

It is remarkable that the classification of Darboux integrable chains is closely connected with the classical 
problem of description of ODE admitting a fundamental system of solutions (following Vessiot-Guldberg-Lie),
for the details one can see \cite{ibr} and the references therein.  
Indeed any $n$-integral defines an ordinary differential equation $F(n,x,t,t_x,...t_{[k]})=p(x)$ for which the corresponding $x$-integral gives a formula $I(n,x,t,t_1,...t_m)=c_n$ allowing one to find a new solution $t_m$ for given set of solutions $t,t_1,...t_{m-1}.$ Iterating this way one finds a solution $t_N=H(t,t_1,...t_{m-1},x,c_1,c_2,...c_k)$ depending on $k$ arbitrary constants. In the case when $I$ does not depend on $x$ explicitly this formula gives general solution in a desired form. Examples are given in the Remark in Section 4.

\section{Characteristic algebras of discrete models}

%\subsection{Definition of the algebra $L_x$}
Let us introduce the characteristic algebras for chain (\ref{dhyp}).
Due to the
requirement of $\frac{\partial f }{\partial t_x} \neq 0$, we can
rewrite (at least locally) chain (\ref{dhyp}) in the  inverse form
$$t_x(n-1, x)=g(x,t(n, x),t(n-1, x),t_x(n, x)).$$
Since $x$-integral $F$  does not depend on the variables $t_{[k]}$, $k\in \NN$, then
the equation $D_xF=0$ becomes $KF=0$, where
\begin{equation}\label{gc1} K = \frac{\partial }{\partial x}+t_x\frac{\partial
}{\partial t} +f\frac{\partial }{\partial t_1 }+g\frac{\partial
}{\partial t_{-1}} +f_1\frac{\partial }{\partial
t_2}+g_{-1}\frac{\partial }{\partial t_{-2} }+\ldots\, .
\end{equation}
Also, $XF=0$,
 with $X=\frac{\partial}{\partial t_x}$. Consider the linear space over the field of locally analytic functions depending on a finite number of dynamical variables spanned by all multiple commutators of $K$ and $X$. This set is closed with respect to three operations: addition, multiplication by a function, and taking the commutator of two elements. It is called characteristic algebra $L_x$ of chain (\ref{dhyp}) in the $x$-direction. Therefore, any vector field from  algebra $L_x$ annulates $F$.
The relation between Darboux integrability of   chain (\ref{dhyp})  and its
characteristic algebra $L_x$ is given by the following  important criterion.
\begin{thm}\label{thm1} (see \cite{HZhP2})
Chain (\ref{dhyp})  admits a
nontrivial $x$-integral if and only if its characteristic algebra $L_x$ is of
finite dimension.
\end{thm}

%\subsection{Definition of the algebra $L_n$}

The
equation $DI=I$, defining an $n$-integral $I$, in an enlarged form becomes
\begin{equation}\label{n-integral}
I(x,n+1, t_1,f,f_{x},...)=I(x,n,t,t_x,t_{xx},...).
\end{equation}
The left hand side contains the variable $t_1$ while the right
hand side does not. Hence we have $D^{-1}\frac{d}{dt_1}DI=0,$ i.e.
the $n$-integral is in the kernel of the operator
$$
Y_1=D^{-1}Y_0D, $$ where
\begin{equation}\label{Y1}
    Y_1=\frac{\partial}{\partial t}+D^{-1}(Y_0f)\frac{\partial}{\partial t_x}+
    D^{-1}Y_0(f_x)\frac{\partial}{\partial t_{xx}}+   D^{-1}Y_0(f_{xx})
    \frac{\partial}{\partial t_{xxx}}+...,
\end{equation}
and \begin{equation} Y_0=\frac{d}{dt_1}\,.
\end{equation}
 One can show that $D^{-j}Y_0D^jI=0$
for any natural $j$.
Direct calculations show that
$$
D^{-j}Y_0D^j=X_{j-1}+Y_j, \qquad j\geq 2,
$$
where
\begin{eqnarray}
&&Y_{j+1}=D^{-1}(Y_jf)\frac{\partial}{\partial t_x}
+D^{-1}Y_j(f_x)\frac{\partial}{\partial t_{xx}}+
D^{-1}Y_j(f_{xx})\frac{\partial}{\partial t_{xxx}}+...,\quad j\geq
1\, , \label{Yj}
\end{eqnarray}
\begin{equation}\label{definitionXj}
X_j=\frac{\partial}{\partial_{t_{-j}}}, \qquad  j\geq
1.\end{equation}
Define by $N^*$ the dimension of the linear space spanned by the operators
$\{Y_j \}_1^{\infty}$. Introduce the linear space over the field of locally analytic functions depending on a finite number of dynamical variables spanned by all multiple commutators of the vector fields from $\{Y_j \}_1^{N^*}\cup\{X_j \}_1^{N^*}$. This linear space is closed with respect to three operations: addition, multiplication by a function, and taking the commutator of two elements. 
It is called characteristic algebra $L_n$ of chain (\ref{dhyp}) in the $n$-direction.

\begin{thm}\label{thm2}(see \cite{hp})
Equation (\ref{dhyp}) admits a nontrivial $n$-integral if and only
 if its characteristic algebra $L_n$ is of
finite dimension.
\end{thm}

\section{On the structure of nontrivial $x$- and $n$- integrals}

We define  the  {\it order of a nontrivial $n$-integral} $I=I(x,n,t,t_x,
\ldots, t_{[k]})$ with $\frac{\partial I}{\partial t_{[k]}}\ne 0$, as the number $k$.

\begin{thm} \label{Thm1}
Assume equation (\ref{dhyp}) admits a nontrivial $n$-integral. Then for any nontrivial  $n$-integral
$I^*(x,n,t,t_x, \ldots, t_{[k]})$ of the smallest order and any $n$-integral $I$ we have
\begin{equation}\label{representationnint}I=\phi(x, I^*, D_xI^*, D^2_xI^*, \ldots),
\end{equation}
where $\phi$ is some function.
\end{thm}
{\it Proof}: Denote by $I^*=I^*(x,n,t, \ldots, t_{[k]})$ an $n$-integral of the smallest order.
Let $I$ be any other $n$-integral, $I=I(x,n, t, \ldots, t_{[r]})$.
Clearly $r\geq k$. Let us introduce new variables $x$, $n$, $t$, $t_x$, $\ldots$, $t_{[k-1]}$, $I^*$,
$D_xI^*$, $\ldots$, $D_x^{r-k}I^*$ instead of the variables  $x$, $n$, $t$, $t_x$, $\ldots$, $t_{[k-1]}$,
$t_{[k]}$, $t_{[k+1]}$, $\ldots$, $t_{[r]}$. Now, $I=I(x, n,  t, t_x, \ldots, t_{[k-1]}, I^*,
D_xI^*, \ldots, D_x^{r-k}I^*)$. We write the power series for function $I$ in the neighborhood of the point
$( (I^{*})_0,
(D_xI^*)_0, \ldots, (D_x^{r-k}I^*)_0)$:
\begin{equation}\label{nintseries}
I=\sum\limits_{i_0, i_1, \ldots, i_{r-k}} E_{i_0, i_1, \ldots,
i_{r-k}}(I^*-(I^{*})_0)^{i_0}(D_xI^*-(D_xI^*)_0)^{i_1}
\ldots (D_x^{r-k}I^*-(D_x^{r-k}I^*)_0)^{i_{r-k}}\, .
\end{equation}
Then
$$
DI=\sum\limits_{i_0, i_1, \ldots, i_{r-k}} DE_{i_0, i_1, \ldots,
i_{r-k}}(DI^*-(I^{*})_0)^{i_0}(DD_xI^*-(D_xI^*)_0)^{i_1}
\ldots (DD_x^{r-k}I^*-(D_x^{r-k}I^*)_0)^{i_{r-k}}\, .
$$
Since $DI=I$, $DD^j_xI^*=D^j_xDI^*=D^j_xI^*$ and the power series representation
for function $I$ is unique, then
$DE_{i_0, i_1, \ldots, i_{r-k}}=E_{i_0, i_1, \ldots, i_{r-k}}$, i.e.   $E_{i_0, i_1,
\ldots, i_{r-k}}(x,n, t, \ldots, t_{[k-1]})$ are all $n$-integrals. Due to the fact that
minimal $n$-integral depends on $x$, $n$, $t$, $\ldots$, $t_{[k]}$,
we conclude that all $E_{i_0, i_1, \ldots, i_{r-k}}(x, n, t, \ldots, t_{[k-1]})$ are trivial
$n$-integrals, i.e. functions depending only on $x$. Now equation (\ref{representationnint})
follows immediately from (\ref{nintseries}).
%$\Box$

We define  the  {\it order of a nontrivial $x$-integral} $F=F(x,n,t_{k},t_{k+1},...t_{m}) $
with $\frac{\partial F}{\partial t_{[m]}}\ne 0$, as the number $m-k$.

\begin{thm} \label{Thm2}
Assume equation (\ref{dhyp}) admits a nontrivial $x$-integral. Then for any nontrivial
$x$-integral $F^*(x,n, t, t_1, \ldots, t_m)$ of the smallest order and any $x$-integral $F$ we have
\begin{equation}\label{representationxint}
F=\xi(n,F^*, DF^*, D^2F^*, \ldots),
\end{equation}
where $\xi$ is some function.
\end{thm}
{\it Proof}:  Denote by $F^*=F^*(x,n, t, t_1, \ldots, t_m)$ an
$x$-integral of the smallest order. Let $F$ be any other
$x$-integral, $F=F(x, n,t, t_1, \ldots, t_l)$. Clearly, $l\geq m$. Let
us introduce new variables $x$, $n$,  $t$, $t_1$, $\ldots$, $t_{m-1}$,
$F^*$, $DF^*$, $\ldots$, $D^{l-m}F^*$ instead of variables $x$, $n$, $t$,
$t_1$, $\ldots$,$t_{m-1}$,  $t_m$, $\ldots$, $t_l$. Now, $F=F(x, n, t, t_1,
\ldots,t_{m-1}, F^*, DF^*, \ldots, D^{l-m}F^*)$. We write the power
series representation of function $F$ in the neighborhood of point
$((F^*)_0, (DF^*)_0, \ldots, (D^{l-m}F^*)_0)$:
\begin{equation}\label{xintseries}
F=\sum\limits_{i_0,i_1, \ldots, i_{l-m}}K_{i_0,i_1, \ldots, i_{l-m}}
(F^*-(F^*)_0)^{i_0}(DF^*-(DF^*)_0)^{i_1}\ldots
(D^{l-m}F^*-(D^{l-m}F^*)_0)^{i_{l-m}}\, .
\end{equation}
Then
$$
D_xF=\sum\limits_{i_0,i_1, \ldots, i_{l-m}}D_x\{K_{i_0,i_1, \ldots, i_{l-m}}\}
(F^*-(F^*)_0)^{i_0}(DF^*-(DF^*)_0)^{i_1}\ldots
(D^{l-m}F^*-(D^{l-m}F^*)_0)^{i_{l-m}}
$$
$$
+\sum\limits_{i_0,i_1, \ldots, i_{l-m}}K_{i_0,i_1, \ldots, i_{l-m}}
D_x\{(F^*-(F^*)_0)^{i_0}(DF^*-(DF^*)_0)^{i_1}\ldots
(D^{l-m}F^*-(D^{l-m}F^*)_0)^{i_{l-m}}\}
$$
Since $D_xD^jF^*=D^jD_xF^*=0$, then 
$$D_x\{(F^*-(F^*)_0)^{i_0}(DF^*-(DF^*)_0)^{i_1}\ldots
(D^{l-m}F^*-(D^{l-m}F^*)_0)^{i_{l-m}}\}=0.$$ Therefore,
$$
0=D_xF=\sum\limits_{i_0,i_1, \ldots, i_{l-m}}D_x\{K_{i_0,i_1, \ldots, i_{l-m}}\}
(F^*-(F^*)_0)^{i_0}(DF^*-(DF^*)_0)^{i_1}\ldots
(D^{l-m}F^*-(D^{l-m}F^*)_0)^{i_{l-m}}\, .
$$
Due to the unique representation of the zero power series we have that
$D_x\{K_{i_0,i_1, \ldots, i_{l-m}}\}=0$, i.e. all $K_{i_0,i_1, \ldots, i_{l-m}}
(x, n, t, \ldots, t_{m-1})$ are
$x$-integrals. Since the minimal nontrivial $x$-integral is of order $m$,
then all $K_{i_0,i_1, \ldots, i_{l-m}}$ are
trivial $x$-integrals, i.e. functions depending on $n$ only.
Now the equation (\ref{representationxint}) follows from (\ref{xintseries}).
%$\Box$

The next two theorems are  discrete versions of  Lemma 1.2 from  \cite{Zhiber}.

\begin{thm} \label{Thm3}
Among all nontrivial $n$-integrals $I^*(x, n, t, t_x, \ldots, t_{[k]})$ of the
smallest order, with $k\geq 2$, there is an $n$-integral $I^0(x, n, t,t_x,
\ldots , t_{[k]})$ such that
\begin{equation}
\label{linearnint}
I^0(x, n,  t, t_x, \ldots, t_{[k]})=a(x, n,t, t_x, \ldots, t_{[k-1]})t_{[k]}+
b(x, n, t, t_x, \ldots, t_{[k-1]})\, .
\end{equation}
\end{thm}
{\it Proof}: Consider nontrivial minimal $n$-integral $I^*(x,n, t, t_x,
\ldots, t_{[k]})$ with $k\geq 2$. Equality $DI^*=I^*$
can be rewritten as
$$
I^*(x, n+1, t_1, f, f_x, \ldots, f_{[k-1]})=I^*(x,n, t,t_x, \ldots, t_{[k]}).
$$
We differentiate both sides of the last equality with respect to $t_{[k]}$:
\begin{equation}\label{**}
\frac{\partial I^*(x,n+1,  t_1, f, \ldots, f_{[k-1]})}{\partial f_{[k-1]}}
\cdot\frac{\partial f_{[k-1]}}{\partial t_{[k]}}=
\frac{\partial I^*(x, n, t, \ldots, t_{[k]})}{\partial t_{[k]}}\, .
\end{equation}
In virtue of $\frac{\partial f_{[j]}}{\partial t_{[j+1]}}=f_{t_x}$, the equation
(\ref{**}) can be rewritten as
\begin{equation}\label{***}
\frac{\partial I^*(x, n+1, t_1, f, \ldots, f_{[k-1]})}{\partial f_{[k-1]}}f_{t_x}=
\frac{\partial I^*(x, n, t, \ldots, t_{[k]})}{\partial t_{[k]}}\, .
\end{equation}
Let us differentiate once more with respect to $t_{[k]}$ both sides of the last equation, we have:
$$
\frac{\partial^2 I^*(x,n+1,  t_1, f, \ldots, f_{[k-1]})}{\partial^2 f_{[k-1]}}f^2_{t_x}=
\frac{\partial^2 I^*(x, n, t, \ldots, t_{[k]})}{\partial t^2_{[k]}}\, ,
$$
or the same,
$$D\left\{ \frac{\partial^2I^*}{\partial t^2_{[k]}}\right\}f^2_{t_x}=
\frac{\partial^2I^*}{\partial t^2_{[k]}}\, ,
$$
where $I^*=I^*(x,n, t,\ldots, t_{[k]})$. It follows from (\ref{***}) that
$$
D\left\{\frac{\partial^2I^*}{\partial t^2_{[k]}}\right\}\left\{
\frac{\partial I^*}{\partial t_{[k]}}\right\}^2=
\frac{\partial^2I^*}{\partial t^2_{[k]}}D\left\{\left(
\frac{\partial I^*}{\partial t_{[k]}}\right)^2\right\}\, ,
$$
or the same, function
$$
J:=\frac{\frac{\partial^2I^*}{\partial t^2_{[k]}}}{\left(
\frac{\partial I^*}{\partial t_{[k]}}\right)^2}
$$ is an $n$-integral, and by Theorem \ref{Thm1}, we have that $J=\phi(x, I^*)$. Therefore,
$$
\frac{\partial^2I^*}{\partial t^2_{[k]}}=\frac{\partial H(x, I^*)}{\partial I^*}\left(
\frac{\partial I^*}{\partial t_{[k]}}\right)^2, \quad \mbox{where} \quad \frac{\partial H}{\partial I^*}=J,
$$
or
$$
\frac{\partial }{\partial t_{[k]}} \left\{ \ln \frac{\partial I^*}{\partial t_{[k]}}-H(x, I^*)\right\}=0.
$$
Hence,  $e^{-H(x, I^*)}
\frac{\partial I^*}{\partial t_{[k]}}=e^g$ for some function $g(x,n, t,t_x, \ldots, t_{[k-1]})$.
Introduce $W$ in such a way that $\frac{\partial W}{\partial I^*}
=e^{-H(x, I^*)}$. Then $\frac{\partial W}{\partial t_{[k]}}=e^g$ and
$W=e^{g(x,n, t, \ldots, t_{[k-1]})}t_{[k]}+l(x,n, t, \ldots, t_{[k-1]})$ is an $n$-integral,
where $l(x,n, t, \ldots, t_{[k-1]})$   is some function.
%$\Box$

\begin{thm} \label{Thm4}
Among all nontrivial $x$-integrals $F^*(x, n, t_{-1}, t, t_1, \ldots, t_m)$ of the smallest order,
with $m\geq 1$,  there is $x$-integral $F^0(x, n, t_{-1}, t, t_{1}, \ldots, t_m)$ such that
\begin{equation}\label{separatexint}
F^0(x,n,  t_{-1}, t, t_{1}, \ldots, t_m)=A(x,n,  t_{-1}, t, \ldots, t_{m-1})+B(x,n,  t, t_1, \ldots, t_m).
\end{equation}
\end{thm}
{\it Proof}: Consider  nontrivial $x$-integral $F^*(x, n, t_{-1}, t, t_1, \ldots, t_m)$ of minimal order.
Since $D_xF^*=0$, then
\begin{equation}\label{O}
\frac{\partial F^*}{\partial x}+g\frac{\partial F^*}{\partial t_{-1}}+
t_x\frac{\partial F^*}{\partial t}+f\frac{\partial F^*}{\partial t_1}+Df
\frac{\partial F^*}{\partial t_2}+\ldots+D^{m-1}f\frac{\partial F^*}{\partial t_m}=0.
\end{equation}
We differentiate both sides of (\ref{O}) with respect to $t_m$ and with respect to
$t_{-1}$ separately and have the following
two equations:
\begin{equation}\label{A}
\{D_x+\frac{\partial }{\partial t_m}(D^{m-1}f)\}\frac{\partial F^*}{\partial t_m}=0,
\end{equation}
\begin{equation}\label{B}
\{D_x+\frac{\partial g }{\partial t_{-1}}\}\frac{\partial F^*}{\partial t_{-1}}=0.
\end{equation}
Let us differentiate (\ref{A}) with respect to $t_{-1}$, we have,
\begin{equation}\label{C}
D_x\frac{\partial^2F^*}{\partial t_m\partial t_{-1}}+\frac{\partial g}{\partial t_{-1}}
\frac{\partial^2F^*}{\partial t_m\partial t_{-1}}+\frac{\partial }{\partial t_m}(D^{m-1}f)
\frac{\partial^2F^*}{\partial t_m\partial t_{-1}}=0\, .
\end{equation}
It follows from (\ref{A}) and (\ref{B}) that $\frac{\partial }{\partial t_m}(D^{m-1}f)=-
\frac{D_xF^*_{t_m}}{F^*_{t_m}}$,
$\frac{\partial g}{\partial t_{-1}}=-\frac{D_x F^*_{t_{-1}}}{F^*_{t_{-1}}}$. Equation (\ref{C}) becomes
$$
D_x\left\{ \ln \frac{F^*_{t_mt_{-1}}}{F^*_{t_m}F^*_{t_{-1}}}\right\}=0.
$$
By Theorem \ref{Thm2} we have, $\frac{F^*_{t_mt_{-1}}}{F^*_{t_m}F^*_{t_{-1}}}=\xi(n, F^*)$, or
$$
\frac{F^*_{t_mt_{-1}}}{F^*_{t_m}}=F^*_{t_{-1}}\xi(n, F^*)=H'(F^*)F^*_{t_{-1}}=
\frac{\partial}{\partial t_{-1}}H(F^*), \quad {\mbox{where }} \quad
\xi(n,F^* )=H'(n, F^*)\, .
$$
Thus,
$\frac{\partial}{\partial t_{-1}}\{ \ln F^*_{t_m} -H(n,F^*)\}=0$, or $e^{-H(n, F^*)}F^*_{t_m}=
C(x,n, t, t_1, \ldots, t_m)$ for some function \\ $C(x,n, t, t_1, \ldots, t_m)$.
Denote by $\tilde{H^*}(n, F)$ such a function that $\tilde{H}'(n, F^*) =e^{-H(n, F^*)}$. Then
$\frac{\partial \tilde{H}(n, F^*)}{\partial t_m}=C(x,n, t,t_1, \ldots, t_m)$. Hence,
$\tilde{H}(n, F^*)=B(x,n, t, t_1, \ldots, t_m)+A(x,n, t_{-1}, t, \ldots, t_{m-1})$.
Since $D_x\tilde{H}(F^*)=\tilde{H}'(n, F^*)D_x(F^*)=0$, then
$\tilde{H}(n, F^*)$ is an $x$-integral in the desired form (\ref{separatexint}).
%$\Box$

\section{Particular case: $t_{1x}=t_x+d(t, t_1)$}

Finiteness of the characteristic algebras $L_x$ and $L_n$ was
used in \cite{HZhP1} and \cite{HZhP2} to classify Darboux integrable
semi-discrete chains of special form
\begin{equation}
\label{main}
t_{1x}=t_x+d(t, t_1)\, .
\end{equation}
The statement of this classification result is given by the next  theorem from \cite{HZhP2}.

\begin{thm}\label{maintheorem}
Chain (\ref{main}) admits nontrivial $x$- and $n$-integrals if and
only if it is one of the kind:\\
(a) \begin{equation}\label{a}t_{1x}=t_x+A(t_1-t),
\end{equation} where $A(t_1-t)$ is given in
implicit form $A(t_1-t)=\frac{d}{d\theta}P(\theta)$,
$t_1-t=P(\theta)$, with $P(\theta)$ being an arbitrary quasi polynomial, i.e. a
function satisfying  an ordinary differential equation
$%\begin{equation}\label{coefficientsmuN}
P^{(N+1)}=\mu_NP^{(N)}+\ldots+\mu_1P'+\mu_0P
%\end{equation}
$with constant coefficients $\mu_k$, $0\leq k\leq N$,\\
(b)\begin{equation}\label{b}t_{1x}=t_x+C_1(t_1^2-t^2)+C_2(t_1-t),\end{equation}
(c)  \begin{equation}\label{c}t_{1x}=t_x+\sqrt{C_3e^{2\alpha
t_1}+C_4e^{\alpha(t_1+t)}+C_3e^{2\alpha t}},
\end{equation}
(d) \begin{equation}\label{d}t_{1x}=t_x +C_5(e^{\alpha
t_1}-e^{\alpha t})+C_6(e^{-\alpha t_1}-e^{-\alpha t}),\end{equation}
\noindent where $\alpha\ne 0$, $C_i$, $1\leq i\leq 6$,  are
arbitrary constants. Moreover, some nontrivial $x$-integrals $F$ and
$n$-integrals $I$ in each of the cases are
\begin{enumerate}
\item[i)] $F=x-\int^{t_1-t}\frac{ds}{A(s)}$, $I=L(D_x)t_x$, where $L(D_x)$ is a differential operator
which annihilates $\frac{d}{d\theta}P(\theta)$ where $D_x\theta=1$.
 \item[ii)]
$F=\frac{(t_3-t_1)(t_2-t)}{(t_3-t_2)(t_1-t)}$, $I=t_x-C_1t^2-C_2t,$
\item[iii)]
$F=arcsinh(ae^{\alpha(t_1-t_2)}+b)+arcsinh(ae^{\alpha(t_1-t)}+b),
\quad a=2C_3/\sqrt{4c_3^2-c_4^2}, \quad b=C_4/\sqrt{4c_3^2-c_4^2}, $
 $I=2t_{xx}-\alpha t_x^2-\alpha C_3
\mathrm{e}^{2\alpha t}$,
\item[iv)]
$F=\frac{(\mathrm{e}^{\alpha t}-\mathrm{e}^{\alpha
t_2})(\mathrm{e}^{\alpha t_1}-\mathrm{e}^{\alpha
t_3})}{(\mathrm{e}^{\alpha  t}-\mathrm{e}^{ \alpha
t_3})(\mathrm{e}^{\alpha t_1}-\mathrm{e}^{\alpha t_2})}$, $I=t_x-C_5
e^{\alpha t}-C_6 e^{-\alpha t}$.
\end{enumerate}
\end{thm}
Note that all the integrals in  Theorem \ref{maintheorem} are given
in their canonical forms (see Theorems \ref{Thm3} and \ref{Thm4}).

\noindent{\bf{Remark}}. In case (c) equation (\ref{c}) is closely connected with
the well-known Steen-Ermakov equation (see \cite{rogers} and the
references therein)
\begin{equation}\label{ermakov}
y''+q(x)y=c_3y^{-3}.
\end{equation}
Indeed for $\alpha=2$ its $n$-integral $2t_{xx}-\alpha t_x^2-\alpha
C_3\mathrm{e}^{2\alpha t}=p(x)$ is reduced to the form
(\ref{ermakov}) by substitution $y=\mathrm{e}^{-t}$. Now it follows
from the $x$-integral $F$  that for arbitrary three solutions
$y(x)$, $z(x)$, $w(x)$ of the Steen-Ermakov equation the following
function
$$ R(y,z,w)=arcsinh({aw^2}{y^{-2}}+b)+arcsinh({az^2}{y^{-2}}+b)$$
does not depend on $x$. Remind that the Riccati equation connected with the cases (b) and (d) has a
similar property: the cross-ratio of its four solutions is a constant.

\subsection{Characteristic algebras $L_x$   for Darboux integrable equations
$t_{1x}=t_x+d(t,t_1)$}

It was proved (see \cite{HZhP1}) that  if equation $t_{1x}=t_x+d(t,t_1)$ admits a
nontrivial $x$-integral, then it admits a nontrivial $x$-integral not depending on $x$.
Introduce new vector fields
$$
\tilde{X}=[X, K]=\sum\limits_{k=-\infty, }^\infty \frac{\partial }{\partial t_k}\,  \qquad 
J:=[\tilde{X}, K]\, .
$$

\subsubsection {Case 1: $t_{1x}=t_x+A(t_1-t)$}

Direct calculations show that the multiplication table for characteristic algebra $L_x$ is the following

\begin{center}
\begin{tabular}{|c||c|c|c|}
\hline
$L_x$ & $X$ & $K$ & $\tilde{X}$    \\
\hline
\hline
$X$   &  0 &   $\tilde{X}$      & 0            \\
\hline
$K$ &  $-\tilde{X}$  & 0  & 0    \\
\hline
 $\tilde{X}$ &  0       &      0   &     0    \\
\hline
\end{tabular}
\end{center}

\subsubsection{Case 2: $t_{1x}=t_x+C_1(t_1^2-t^2)+C_2(t_1-t)$}

Direct calculations show that
$$J=2C_1\sum\limits_{k=-\infty, k\ne 0}^\infty (t_k-t) \frac{\partial }{\partial t_k}$$
and
$$
[J, K]=2C_1^2\sum\limits_{k=-\infty, k\ne 0}^\infty (t_k-t)^2 \frac{\partial }{\partial t_k}
=2C_1(K-t_x \tilde{X})-(2C_1t+C_2)J\, \,$$
and the multiplication table for characteristic algebra $L_x$ is the following

\begin{center}
\begin{tabular}{|c||c|c|c|c|}
\hline
$L_x$ & $X$ & $K$ & $\tilde{X}$ & $J$\\
\hline
\hline
$X$ & 0 & $\tilde{X}$ & 0 & 0\\
\hline
$K$ & $-\tilde{X}$ & 0 & $-J$ & $-2C_1(K-t_x \tilde{X})+(2C_1t+C_2)J$\\
\hline
$\tilde{X}$ & 0 & $J$ & 0 & 0\\
\hline
$J$ & 0 & $2C_1(K-t_x \tilde{X})-(2C_1t+C_2)J$ & 0& 0\\
\hline
\end{tabular}
\end{center}

\subsubsection{Case 3: $t_{1x}=t_x +\sqrt{C_3e^{2\alpha t_1}+C_4e^{\alpha (t_1+t)}+C_3e^{2\alpha t}}$}

Direct calculations show that $[\tilde{X}, K]=\alpha K-\alpha t_x \tilde{X}$, and the multiplication
table for characteristic algebra $L_x$ is the following

\begin{center}
\begin{tabular}{|c||c|c|c|}
\hline
$L_x$ & $X$ & $K$ & $\tilde{X}$    \\
\hline
\hline
$X$   &  0 &   $\tilde{X}$      & 0            \\
\hline
$K$ &  $-\tilde{X}$  & 0  &  $-\alpha K+\alpha t_x \tilde{X}$  \\
\hline
 $\tilde{X}$ &  0       &      $\alpha K-\alpha t_x \tilde{X}$  &     0    \\
\hline
\end{tabular}
\end{center}

\subsubsection{Case 4: $t_{1x}=t_x+C_5(e^{\alpha t_1}-e^{\alpha t})
+C_6(e^{-\alpha t_1}-e^{-\alpha t})$}

Direct calculations show that
$$
J=\alpha \sum\limits_{k=-\infty, k\ne 0}^\infty \{ C_5(e^{\alpha t_k}-e^{\alpha t})-
C_6(e^{-\alpha t_k}-e^{-\alpha t})\}
\frac{\partial }{\partial t_k}
$$
and
$$
[J, K]=2C_5C_6\alpha^2 \sum\limits_{k=-\infty, k\ne 0}^\infty \{  e^{\alpha (t-t_k)}+
e^{\alpha (t_k-t)}-2\}
\frac{\partial }{\partial t_k}
$$
$$
= \alpha^2(C_5e^{\alpha t}+C_6e^{-\alpha t})(K-t_x \tilde{X})+\alpha (C_6e^{-\alpha t}
-C_5e^{\alpha t})J.
$$
Denote by
$$
\beta_1=\alpha^2(C_5e^{\alpha t}+C_6e^{-\alpha t}), \qquad \beta_2=\alpha (C_6e^{-\alpha t}
-C_5e^{\alpha t})\, .
$$
The multiplication table for characteristic algebra $L_x$ is

\begin{center}
\begin{tabular}{|c||c|c|c|c|}
\hline
$L_x$ & $X$ & $K$ & $\tilde{X}$ & $J$\\
\hline
\hline
$X$ & 0 & $\tilde{X}$ & 0 & 0\\
\hline
$K$ & $-\tilde{X}$ & 0 & $-J$ &$-\beta_1(K-t_x \tilde{X})-\beta_2 J $\\
\hline
$\tilde{X}$ & 0 & $J$ & 0 &$\alpha^2K-\alpha^2\tilde{X}$\\
\hline
$J$ & 0 & $\beta_1(K-t_x \tilde{X})+\beta_2 J$ & $\alpha^2\tilde{X}-\alpha^2K$& 0\\
\hline
\end{tabular}
\end{center}

\subsection{Characteristic algebras $L_n$   for Darboux integrable equation
$t_{1x}=t_x+d(t,t_1)$}

\subsubsection {Case 1: $t_{1x}=t_x+A(t_1-t)$}

Characteristic algebra $L_n$ is generated only by two vector fields $X_1$ and $Y_1$, and can be of
any finite dimension.
If $A(t_1-t)=t_1-t+c$, where $c$ is some constant, then characteristic algebra $L_n$ is trivial,
consisting of $X_1$ and $Y_1$ only, with commutativity relation $[X_1, Y_1]=0$. If $A(t_1-t)\ne t_1-t+c$,
one can choose a basis in $L_n$
consisting of $W=\frac{\partial }{\partial \theta}$, $Z=\sum_{k=0}^{k=\infty}
D_x^kp(\theta)\partial/\partial t_{[k]}$, with $\theta=x+\alpha_n$,
$C_1=[W, Z]$, $C_{k+1}=[W, C_k]$, $1\leq k\leq N-1$. Its multiplication table for $L_n$ is the following

\begin{center}
\begin{tabular}{|c||c|c|c|c|c|c|c|c|c|}
\hline
$L_n$ & $W$ &  $Z$   & $C_1$ & $C_2$  & $\ldots$& $C_k$     & $\ldots$ & $C_{N-1}$ & $C_N$  \\
\hline
\hline
$W$   &  0      & $C_1$ & $C_2$ &  $C_3$     &$\ldots$ & $C_{k+1}$ & $\ldots$ &  $C_N$       &$K$ \\
\hline
$Z$  &  $-C_1$  & 0      & 0    & 0      & $\ldots$ &   0        & $\ldots$ &0   &0  \\
\hline
$C_1$ & $-C_2$  & 0     & 0      & 0       &$\ldots$ & 0          & $\ldots$ &0&0  \\
\hline
 $\vdots$ & $\vdots$ &  $\vdots$ &  $\vdots$ &  $\vdots$ &  $\vdots$ &  $\vdots$ &  $\vdots$ &
 $\vdots$ &  $\vdots$            \\
 \hline
 $C_N$ &  $-K$  & 0  & 0  & 0 &$\ldots$ & 0 & $\ldots$& 0&0  \\
\hline
\end{tabular}
\end{center}
where $K=\mu_0 Z+\mu_1C_1+\ldots + \mu_N C_N$.

\subsubsection{Cases 2 and 4: $t_{1x}=t_x+C_1(t_1^2-t^2)+C_2(t_1-t)$ and $t_{1x}=t_x+
C_5(e^{\alpha t_1}-e^{\alpha t})
+C_6(e^{-\alpha t_1}-e^{-\alpha t})$}

In both cases characteristic algebra $L_n$ is trivial, consisting of $X_1$ and $Y_1$ only, with
commutativity relation $[X_1, Y_1]=0$.

\subsubsection{Case 3: $t_{1x}=t_x +\sqrt{C_3e^{2\alpha t_1}+C_4e^{\alpha (t_1+t)}+C_3e^{2\alpha t}}$}

Denote by $\tilde{X}_1=A(\tau_{-1})e^{-\alpha\tau_{-1}}\frac{\partial}{\partial \tau_{-1}}$
and $\tilde{Y}_1=A(\tau_{-1})Y_1$, $C_2=[\tilde{X}_1, \tilde{Y}_1]$. Direct calculations show
that  the  multiplication table for algebra $L_n$ is the following
\begin{center}
\begin{tabular}{|c||c|c|c|}
\hline
$L_n$ & $\tilde{X}_1$ & $\tilde{Y}_1$ & $C_2 $   \\
\hline
\hline
$\tilde{X}_1$   &  0 &   $C_2$      &  $\alpha^2C_3\tilde{Y}_1+C_4/(2C_3)\tilde{X}_1  $        \\
\hline
$\tilde{Y}_1$&  $-C_2$  & 0  &$ K$   \\
\hline
 $C_2$ & $-\alpha^2C_3\tilde{Y}_1-C_4/(2C_3)\tilde{X}_1  $        &   $-K$      &     0    \\
\hline
\end{tabular}
\end{center}
where $K=-(\alpha^2C_4/2)\tilde{Y}_1+(2\alpha^2C_4e^{\alpha\tau_{-1}}-
\alpha^2C_3)\tilde{X}_1$.

\subsection{Explicit solutions for Darbour integrable chains from Theorem \ref{maintheorem}}

Below we find explicit solutions for Darboux integrable chains of
special form (\ref{main}).

\begin{thm}\label{TheoremExplicit}
(a) The explicit solution of equation (\ref{a})  is \begin{equation}\label{aexp}
t(n,x)=t(0,x)+\sum^{n-1}_{j=0} R(x+P_j),
\end{equation}
where $t(0,x)$ and $P_j$ are arbitrary functions of $x$ and $j$
respectively, and $A(\tau)=R'(\theta)$, $t_1-t=R(\theta)$.\\
(b) The explicit solution of equation (\ref{b}) is \begin{equation}\label{bexp}
t(n,x)=\frac{1}{C_1}\left(\frac{\psi_{xx}}{2\psi_x}-\frac{\psi_x}{P_n+\psi} \right)-\frac{C_2}{2C_1},
\end{equation}
where $\psi=\psi(x)$ is an arbitrary function depending on $x$ and $P_n$
is an arbitrary function depending on $n$ only.  \\
(c) The explicit solution $t(n,x)$ of equation (\ref{c}) satisfies \begin{equation}\label{cexp}
e^{\alpha t(n,x)}=\frac{\mu'(x)(R_1(P_n-P_{n+1}))}{0.25\alpha^2(\mu(x)+(P_n+P_{n+1})+
R_3(P_n-P_{n+1}))^2-C_3(R_1(P_n-P_{n+1}))^2}\, ,
\end{equation}
where $R_1=2\alpha/\sqrt{2C_3+C_4}$, $R_3=\sqrt{2C_3-C_4}/\sqrt{2C_3+C_4}$, $\mu$ and $P_n$ are
arbitrary functions depending respectively on variables $x$ and $n$.  \\
(d) Equation (\ref{d}) does not admit any explicit formula for general solution of the form
\begin{equation}\label{2case4}
t=H(x, \psi(x), \psi'(x), \ldots, \psi^{(k)}(x), P_n, P_{n+1}, \ldots, P_{n+m}).
\end{equation}
However, equation (\ref{d}) admits general solution in more complicated form
\begin{equation}\label{case4}
e^{\alpha t(n,x)}=\frac{1}{\alpha C_5}\left(\frac{\psi_{xx}}{2\psi_x}-\frac{\psi_x}{P_n+\psi}\right)+\frac{1}{\alpha C_5} w\, ,
\end{equation}
 where the nonlocal variable $w=w(x)$ is a solution of the first order ODE
$$
w_x'+w^2-\alpha^2C_5C_6=-\left(\frac{\psi_{xx}}{2\psi_x}\right)_x+
\left(\frac{\psi_{xx}}{2\psi_x}\right)^2.
$$
\end{thm}

\noindent{\it{Proof}}: In a trivial case (a) the explicit solution
was described in \cite{HZhP2}.\\
In case $(b)$, the equation (\ref{b}) has an $n$-integral
$I=t_x-C_1t^2-C_2t$. Since $DI=I$ then we have the following Riccati
equation  $t_x-C_1t^2-C_2t=C(x)$ to solve and obtain the explicit
solution (\ref{bexp}).

In case $(c)$, to find the explicit  solution of equation (\ref{c})
we look for the Cole-Hopf type substitution $t=H(v,v_1,v_x)$ that reduces the equation
to the semi-discrete D'Alembert equation $v_{1x}=v_x$ for which the solution is  $v=\mu(x)+P_n$.
Let us find function $H(v,v_1,v_x)$. Since  $v_{1x}=v_x$, then $t_1=H(v_1,v_2,v_x)=:\bar{H}$ and
$t_x=v_{xx}H_{v_x}+v_x(H_v+H_{v_1})$, $t_{1x}=v_{xx}\bar{H}_{v_x}+v_x(\bar{H}_{v_1}+\bar{H}_{v_2})$.
In new variables equation
(\ref{c}) becomes
\begin{equation}\label{3case3}
v_{xx}\{\bar{H}_{v_x}-H_{v_x}\}=v_x(H_v+H_{v_1}-\bar{H}_{v_1}-\bar{H}_{v_2})+
\sqrt{C_3e^{2\alpha \bar{H}}+C_4e^{\alpha(H+\bar{H})}+C_3e^{2\alpha H}}\,.
\end{equation}
The right side of (\ref{3case3}) does not depend on $v_{xx}$, but the left side does unless
$\bar{H}_{v_x}=H_{v_x}$. It implies that $H(v,v_1,v_x)=\psi(v_x)+A(v,v_1)$, where $\psi$ is a
function of one variable $v_x$ and $A$ is a function depending on $v$ and $v_1$. Now,
$t=H(v,v_1,v_x)=\psi(v_x)+A(v,v_1)$,
$t_1=\psi(v_x)+A(v_1,v_2)=:\psi(v_x)+\bar{A}$, and equality (\ref{3case3}) becomes
\begin{equation}\label{4case3}
(\bar{A}_{v_1}+\bar{A}_{v_2}-A_v-A_{v_1})v_x=e^{\alpha \psi(v_x)}
\sqrt{C_3e^{2\alpha \bar{A}}+C_4e^{\alpha(A+\bar{A})}+C_3e^{2\alpha A}}
\end{equation}
that shows $e^{\alpha \psi(v_x)}=Rv_x$, where $R$ is some constant. We have,
\begin{equation}\label{2case3}
e^{\alpha H(v,v_1,v_x)}=Rv_xe^{\alpha A(v,v_1)}\, .
\end{equation}
Let us find function $A(v,v_1)$. In variables $v_k$, $v_{[k]}$ an $n$-integral
$I=2t_{xx}-\alpha t_x^2-\alpha C_3e^{2\alpha t}
=q(x)$  of equation (\ref{c}) becomes
$$
I=\left( \frac{2v_{xxx}}{\alpha v_x}-\frac{3v^2_{xx}}{\alpha v^2_x}\right)+
v_x^2(2A_{vv}+4A_{vv_1}+2A_{v_1v_1}-\alpha(A_v+A_{v_1})^2
-\alpha C_3R^2e^{2\alpha A})
$$
$$
=:s(v_x,v_{xx}, v_{xxx})+v_x^2p(v,v_1)=q(x)\, .
$$
One can see
$$p(v,v_1)=2A_{vv}+4A_{vv_1}+2A_{v_1v_1}-\alpha(A_v+A_{v_1})^2
-\alpha C_3R^2e^{2\alpha A}=0,$$
that can be rewritten in variables $\xi=(v+v_1)/2$ and $\eta=(v-v_1)/2$ in the following form
$$
2A_{\xi\xi}-\alpha (A_\xi)^2-\alpha C_3R^2e^{2\alpha A}=0
$$
that implies
\begin{equation}\label{5case3}
e^{-\alpha A(\xi,\eta)}=\frac{\alpha^2}{4}C_1(\eta)(\xi+C_2(\eta))^2-\frac{C_3R^2}{C_1(\eta)},
\end{equation}
where $C_1(\eta)$ and $C_2(\eta)$ are some functions depending on $\eta$ only.
Equation (\ref{4case3}) implies not only $e^{\alpha \psi(v_x)}=Rv_x$, but also
$$
\bar{A}_{v_1}+\bar{A}_{v_2}-A_v-A_{v_1}=R
\sqrt{C_3e^{2\alpha \bar{A}}+C_4e^{\alpha(A+\bar{A})}+C_3e^{2\alpha A}}
$$
that in variables $\xi$ and $\eta$ becomes
\begin{equation}\label{7case3}
\bar{A}_{\xi_1}-A_\xi=R\sqrt{C_3e^{2\alpha \bar{A}}+C_4e^{\alpha(A+\bar{A})}+C_3e^{2\alpha A}}\, .
\end{equation}
We find function $A$ from (\ref{5case3}) and substitute it into (\ref{7case3}),
remembering that $\xi_1=\xi-\eta-\eta_1$. We have,
$$
\frac{4}{\alpha^2C_3R^2}\left\{-\frac{\xi-\eta-\eta_1+C_2(\eta_1)}{
(\xi-\eta-\eta_1+C_2(\eta_1))^2-4\alpha^{-2}C_3R^2}+
\frac{\xi+C_2(\eta)}{(\xi+C_2(\eta))^2-4\alpha^{-2}C_3R^2}\right\}^2
$$
$$
=\left(\frac{4\alpha^{-2}C_1^{-1}(\eta_1)}{(\xi-\eta-\eta_1+C_2(\eta_1))^2-
4\alpha^{-2}C_3R^2C_1^{-2}(\eta_1)}\right)^2 +
\left( \frac{4\alpha^{-2}C_1^{-1}(\eta)}{(\xi+C_2(\eta))^2-4\alpha^{-2}C_3R^2C_1^{-2}(\eta)}\right)^2
$$
$$
+\frac{16\alpha^{-4}C_4C_3^{-1}C_1^{-1}(\eta)C_1^{-1}(\eta_1)}{((\xi+C_2(\eta))^2-
4\alpha^{-2}C_3R^2C_1^{-2}(\eta))
((\xi-\eta-\eta_1+C_2(\eta_1))^2-4\alpha^{-2}C_3R^2C_1^{-2}(\eta_1))}\, ,
$$
or, equivalently, 
\begin{equation}\label{6case3}
(C_2(\eta)+\eta-(C_2(\eta_1)-\eta_1))^2-4\alpha^{-2}C_3R^2(C_1^{-2}(\eta)+C_1^{-2}(\eta_1))\, 
-4\alpha^{-2}C_4R^2C_1^{-1}(\eta)C_1^{-1}(\eta_1)=0.\end{equation}
 To find $C_1(\eta)$ and $C_2(\eta)$, let us differentiate both sides of (\ref{6case3})
with respect to $\eta$
and then with respect to $\eta_1$. We have,
$$
\frac{-2C_4R^2C_1'(\eta)}{C_1^2(\eta)(C_2'(\eta)+1)}=\frac{(C_2'(\eta_1)-1)C_1^2(\eta_1)}{C_1'(\eta_1)}\, .
$$
We use the fact that the left side of the last equation depends on $\eta$ only, but the right side depends only on
$\eta_1$ and obtain
\begin{equation}\label{8case3}
C_1(\eta)=\frac{1}{R_1\eta+R_2}, \qquad C_2(\eta)=R_3\eta+R_4,
\end{equation}
where $R_1=2(D+2\alpha^{-2}C_4R^2D^{-1})^{-1}$, $R_3=(-D+2\alpha^{-2}C_4R^2D^{-1})(D+2\alpha^{-2}C_4R^2D^{-1})^{-1}$, and $D$, $R_2$, $R_4$ are some constants.
Combining formulas (\ref{2case3}), (\ref{5case3}) and (\ref{8case3}) we see that
\begin{equation}\label{9case3}
e^{\alpha t}=\frac{\mu'(x)R(R_1(P_n-P_{n+1})+R_2)}{0.25\alpha^2(\mu(x)+(P_n+P_{n+1})+
R_3(P_n-P_{n+1})+R_4)^2-C_3(R_1(P_n-P_{n+1})+R_2)^2}\, .
\end{equation}
Without loss of generality, we may assume $R=1$ and $R_4=0$. Substitution
of (\ref{9case3}) into (\ref{c}) shows that $R_2=0$, 
$R_1= 2\alpha/\sqrt{2C_3+C_4}$ and $R_3=\sqrt{2C_3-C_4}/\sqrt{2C_3+C_4}$.

Let us study case $(d)$. For the sake of convenience we set $C_5=C_6=\alpha=1$.
Suppose that there exists function $H$ such that for any choice of functions  $\psi(x)$ and $P_n$
function (\ref{2case4}) solves (\ref{d}).
Consider the variables $\psi$, $\psi'$, $\ldots$, $\psi^{(j)}$, $\ldots$,
$P_n$, $P_{n\pm 1}$, $\ldots$, $P_{n\pm k}$, $\ldots$ as new dynamical variables. Substitution of  $t=H$,
$t_1=H(x,\psi(x), \psi'(x), \ldots, \psi^{(k)}, P_{n+1}, P_{n+2}, \ldots, P_{n+m+1})=:\bar{H}, $
$t_x=H_x+\psi'H_{\psi}+\psi''H_{\psi'}+\ldots+\psi^{(k+1)}H_{\psi^{(k)}},$
$t_{1x}=\bar{H}_x+\psi'\bar{H}_{\psi}+\psi''\bar{H}_{\psi'}+\ldots+\psi^{(k+1)}\bar{H}_{\psi^{(k)}}$
into (\ref{d}) yields
\begin{equation}\label{3case4}
\bar{H}_x+\psi'\bar{H}_{\psi}+\psi''\bar{H}_{\psi'}+\ldots+\psi^{(k+1)}\bar{H}_{\psi^{(k)}}=H_x+
\psi'H_{\psi}+\psi''H_{\psi'}+\ldots+\psi^{(k+1)}H_{\psi^{(k)}}
\end{equation}
$$
+e^{\bar{H}}+e^{-\bar{H}}-e^H-e^{-H}.
$$
Evidently, $H_{\psi^{(k)}}=\bar{H}_{\psi^{(k)}}$ and  consequently $H$ can be represented as
\begin{equation}\label{H}
H=h(x, \psi, \psi', \ldots, \psi^{(k)})+r(x,\psi, \psi', \ldots, \psi^{(k-1)}, P_n, P_{n+1}, \ldots, P_{n+m}).
\end{equation}
Substitute $H$ found in (\ref{H}) in the relation $t_x-2\cosh t=C(x)$ obtained from the $n$-integral
for (\ref{d}). We have,
\begin{equation}\label{4case4}h_x+r_x+
(h_\psi+r_\psi)\psi'+(h_{\psi'}+r_{\psi'})\psi''+\ldots
\end{equation}
$$
(h_{\psi^{(k-1)}}+r_{\psi^{(k-1)}})\psi^{(k)}
+h_{\psi^{(k)}}\psi^{(k+1)}-2\cosh(h+r)=C(x).
$$
Differentiate it with respect to $P_{n+m}:=z$
\begin{equation}
r_{xz}+
r_{\psi z}\psi' + r_{\psi' z}\psi''+\dots + r_{\psi^{k-1} z}\psi^{(k)}-2r_z\cosh(h+r) =0.
\end{equation}
We find $h+r$ from the last equation
\begin{equation}\label{5case4}
h+r=arccosh \left( \frac{r_{xz}+r_{\psi z} \psi' + \dots +r_{\psi^{k-1} z} \psi^{(k)}}{2r_z}\right)\, .
\end{equation}
Denote $r_{xz}+r_{\psi z} \psi' + \dots +r_{\psi^{k-2} z}
\psi^{(k-1)}=A(x,\psi, \psi', \psi^{(k-1)}, P_n,\dots,P_{n+m})$.
Differentiate (\ref{5case4}) with respect to $z=P_{n+m}$ and set
\begin{equation}
r_z= \frac{\left(\frac{A+r_{\psi^{(k-1)} z} \psi^{(k)}} {2r_z}\right)_z}{
\sqrt{\left(\frac{A+r_{\psi^{(k-1)} z} \psi^{(k)}} {2r_z}\right)^2-1}},
\end{equation}
where $A$ and $r$ do not depend on $\psi^{(k)}$. Let us denote $p:=r_z$, $y:=\psi^{(k-1)}$ and $\xi:=\psi^{(k)}$
then
\begin{equation}
p^2\left( \frac{A+p_y\xi}{2p}\right)^2-p^2=\left( \frac{A+p_y\xi}{2p}\right)_z=\left(
\frac{(A_z+p_{yz}\xi)p-p_z(A+p_y\xi)}{2p^2}\right)^2\,
\end{equation}
or
\begin{equation}\label{6case4}
p^4(A+p_y\xi)^2-4p^6=((A_zp-p_zA)+(p_{yz}p-p_yp_z)\xi)^2  .
\end{equation}
Comparison of the coefficients in (\ref{6case4}) gives rise to three
equalities:  $p^4p_y^2=(p_{yz}p-p_yp_z)^2$,  $p^4Ap_y
=(A_zp-p_zA)(p_{yz}p-p_yp_z)$
and $p^4A^2-4p^6=(A_zp-p_zA)^2$,
that are consistent only if  $p=r_z=r_{P_{n+m}}=0$. This condition
$H_{P_{n+m}}=0$ contradicts to our assumption that $H$ essentially
depends on $P_{n+m}$. (Note that if $H_{P_k}=0$ for all $k\in \ZZ$,
then we have a trivial  solution $t=H(x)$ for equation (\ref{d}).)
Therefore, in case (d) the solution can not be represented in form
(\ref{2case4}).

One can use $n$-integral  $I=t_x-C_5e^{\alpha t}-C_6e^{-\alpha t}$, solve the  equation $I=C(x)$
which with the help of the substitution $u=e^{\alpha t}$ can be brought to Riccati equation, and
see that equation (\ref{d})  admits a general solution in more complicated form (\ref{case4}).

\section{Conclusions}

Darboux integrable semi-discrete chains are studied. Structures of their integrals are described. It is proved that if the chain admits an $n$-integral of order $k$ then it admits also an $n$-integral linearly depending on the highest order variable $t_{[k]}$. Similarly, 
if the chain admits an $x$-integral $F(x,t_{-k},t_{-k+1},...t_{m})$, then there is an $x$-integral $F^0(x,t_{-k},t_{-k+1},...t_{m})$
solving the equation 
$$\frac{\partial^2F^0}{\partial t_{-k}\partial t_{m}}=0.$$

The previously found list of Darboux integrable chains of a particular form $t_{1x}=t_x+d(t,t_1)$ is studied in details. The tables of multiplication for the corresponding characteristic algebras are given, explicit formulas for general solutions are constructed. 

The problem of complete classification of Darboux integrable chains (\ref{dhyp}) is still open. Another important open problem is connected with systems of discrete equations: find Darboux integrable discrete versions of the exponential type hyperbolic systems corresponding to the Cartan matrices of semi-simple Lie algebras.

\section*{Acknowledgments}
This work is partially supported by the Scientific and
Technological Research Council of Turkey (T\"{U}B{\.{I}}TAK)  grant $\# $209 T 062,
Russian Foundation for Basic
Research (RFBR) (grants $\#$ 09-01-92431KE-a, $\#$
08-01-00440-a, $\#$ 10-01-91222-CT-a and $\#$ 10-01-00088-a), and MK-8247.2010.1.

\end{document}